# Beyond Stereotypes: Exploring How Minority College Students Experience Stigma on Reddit

Chaeeun Han[1], Sangpil Youm[2], Hojeong Yoo[3], and Sou Hyun Jang[4]

[1]Pennsylvania State University, University Park, PA, USA
[2]University of Florida, Gainesville, FL, USA
[3]University of Minnesota, Minneapolis, MN, USA
[4]Korea University, Seoul, Korea

*Keywords:* Stereotype Detection, Stigma Detection, Social Media Analysis, Keyword Networks, Online Discourse

## Extended Abstract[1]

Higher education in the U.S. has diversified in many aspects, including gender, race/ethnicity, nativity, and religion (Hirschman, 2016). Minority students in college often face unique challenges shaped by their intersecting identities, such as gender, race, religion, and institutional affiliation (Denson, Bowman, Ovenden, Culver, & Holmes, 2021). These challenges include experiences of stereotyping, exclusion, and systemic inequalities, all of which can negatively impact their academic performance, mental health, and sense of belonging (Brannon & Lin, 2021). A research extensively explores the effects of minority status on educational outcomes and social experiences, highlighting disparities in access, retention, and achievement compared to non-minority students (Stephens, Fryberg, Markus, Johnson, & Covarrubias, 2012).

While significant strides have been made in understanding the experiences of individual minority groups, the stigma-related processes shaping these experiences remain unexplored. Existing studies on college minority students' experiences with stereotyping and discrimination predominantly rely on conventional methods, such as small-scale surveys or interviews with a limited number of participants. To address these limitations, the current study implements a novel approach by capturing both large-scale patterns and nuances in the interactions among minority college students. Active participation by college students in online communities (Li, Xie, Chiu, & Ho, 2023) reveals new insights through the analysis of Reddit data, reflecting their lived experiences and interactions. This approach addresses the limitation of dataset scale and offers a novel method to uncover insights that conventional methods may overlook.

The current study aims to achieve two primary objectives. First, drawing on Link and Phelan's (Link & Phelan, 2001) conceptualization of stigma as a process, our approach examines how the key components of stigma—labeling, stereotyping, separation, status loss, and discrimination—emerge across different minority statuses, such as gender, race, religion, and profession. Second, it explores how these components interact and evolve when multiple minority statuses intersect, illustrating the compounded stigmatization faced by individuals with overlapping identities. By analyzing both singular and intersecting statuses, the study aims to deepen our understanding of how stigma operates in diverse contexts.

---

[1]Note: This paper includes examples of offensive language intended for research analysis.

To achieve this, we develop a novel framework that integrates advanced natural language processing techniques with network analysis and quantitative measures of semantic similarity to capture the multiple dimensions of stigma–labeling, stereotyping, separation, and status loss and discrimination–as defined by Link and Phelan (Link & Phelan, 2001) as described in Figure 1. Our approach uses Stereotype-BERT (Zekun, Bulathwela, & Koshiyama, 2023) and sentence-level contextual embeddings to assess semantic distances with reference sentences and systematically detect, quantify, and visualize these facets in online discourse. Additionally, constructing an undirected keyword network from Reddit posts enables the examination of how topics and linguistic cues co-occur across intersecting minority statuses, providing a comprehensive, data-driven perspective on stigmatization among minority college students.

Findings from our analysis classify online posts into four primary categories—Gender, Race, Profession, and Religion—and six intersectional combinations (e.g., Gender & Race). As shown in Table 1, gender posts often discuss everyday campus issues like roommate dynamics, whereas race posts focus on experiences involving non-American ethnicity, such as encounters with Chinese professors or challenges faced by international students. Profession posts scrutinize academic credibility and faculty performance, and religion posts relate to campus religious activities and Christian institutions. Posts with overlapping labels capture more complex dynamics—for example, a gender & profession post may critique preferential grading for "pretty girls," and a race & religion post might explore the struggles of an openly gay student at a conservative Christian university.

Quantitative measures, as shown in Figure 2, reveal distinct patterns in the stigma process. Profession posts exhibit the highest levels of early-stage stereotyping and separation, while gender posts show moderate stereotyping but higher separation. Conversely, race posts face higher discrimination rates despite lower stereotyping. This shows that, although professional identities face early negative stereotyping, racial identities are more vulnerable to later consequences such as status loss and discrimination. Intersectional analyses indicate that overlapping minority statuses, particularly those involving religion, intensify stigma, highlighting compounded vulnerabilities and a more severe progression of exclusion.

Building on these findings, we construct keyword networks to visualize semantic representations and interrelations among the different dimensions of stigma. In the stereotype keyword network (see Figure 3), terms spanning all four categories (e.g., "campus," "university") serve as bridges linking category-specific expressions—such as "friend" and "freshman" for gender or "Christian school" and "god" for religion—and overlapping labels reveal shared concepts like "roommate" or "exam" that underscore common campus experiences.

In the separation keyword network, as shown in Figure 4, keywords such as "fake id," "communal residence," and "crime semester" highlight both physical and perceived isolation. When multiple stigmatized identities intersect, the language intensifies with emotionally charged terms like "depression," "struggle immensely," and even explicit expressions of anger, reflecting compounded exclusion and self-stigmatization.

Similarly, the discrimination & status Loss network, as shown in Figure 5, uncovers recurring terms like "rejection" and "fraud," capturing students' narratives of both systemic bias and personal marginalization. These networks demonstrate that each stigma dimension has distinct lexical markers, but overlapping minority statuses amplify these effects, leading to deeper and more pervasive exclusion.

Future research should broaden the exploration of stigma by examining both single and intersecting minority statuses in contexts beyond education, such as workplaces, healthcare, and social services. Investigating how historical, cultural, and political factors influence stigma in different societies may deepen our understanding of its diverse manifestations. Adopting longitudinal designs to track the evolution of stigma—from early stereotyping to discrimination and status loss—helps inform more effective interventions and policy recommendations.

# References

Brannon, T. N., & Lin, A. (2021). "pride and prejudice" pathways to belonging: Implications for inclusive diversity practices within mainstream institutions. *American Psychologist*, *76*(3), 488.

Denson, N., Bowman, N. A., Ovenden, G., Culver, K., & Holmes, J. M. (2021). Do diversity courses improve college student outcomes? a meta-analysis. *Journal of Diversity in Higher Education*, *14*(4), 544.

Hirschman, C. (2016). *From high school to college: Gender, immigrant generation, and race-ethnicity*. Russell Sage Foundation.

Li, S., Xie, Z., Chiu, D. K., & Ho, K. K. (2023). Sentiment analysis and topic modeling regarding online classes on the reddit platform: educators versus learners. *Applied Sciences*, *13*(4), 2250.

Link, B. G., & Phelan, J. C. (2001). Conceptualizing stigma. *Annual review of Sociology*, *27*(1), 363–385.

Stephens, N. M., Fryberg, S. A., Markus, H. R., Johnson, C. S., & Covarrubias, R. (2012). Unseen disadvantage: how american universities' focus on independence undermines the academic performance of first-generation college students. *Journal of personality and social psychology*, *102*(6), 1178.

Zekun, W., Bulathwela, S., & Koshiyama, A. S. (2023). Towards auditing large language models: Improving text-based stereotype detection. *arXiv preprint arXiv:2311.14126*.

# Figures and Tables

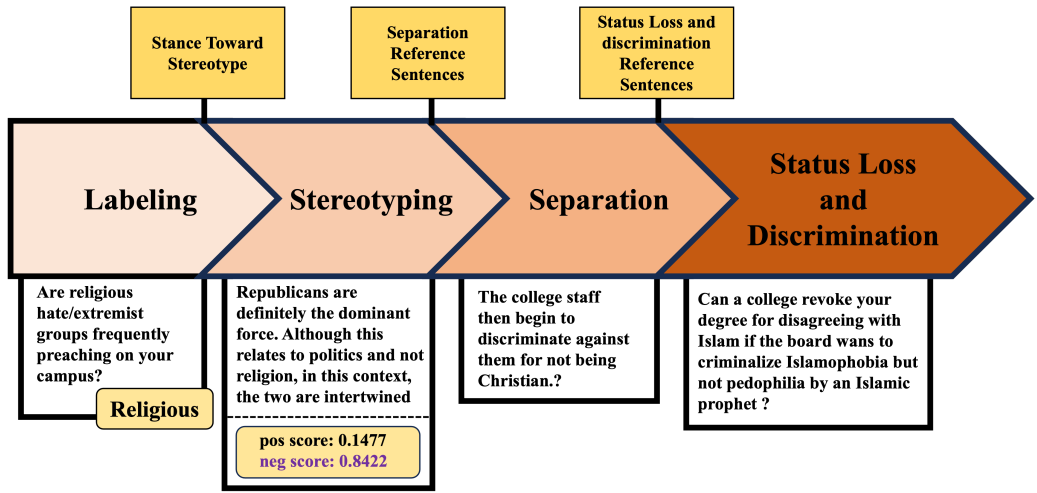

Figure 1: Flow of Examining State Stigma.

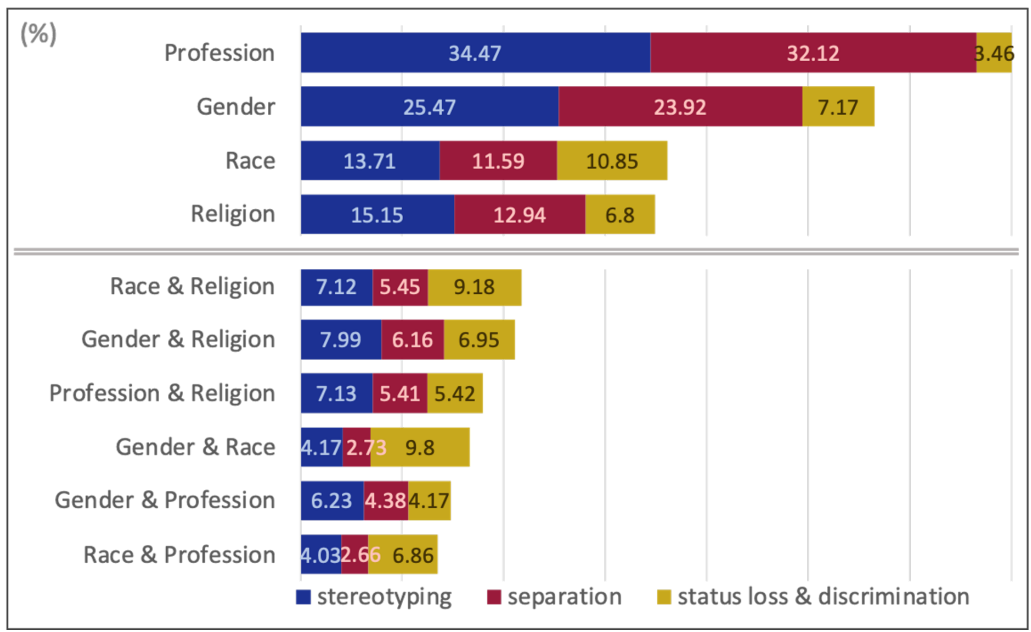

Figure 2: Comparison of Stereotyping, Separation, and Status Loss & Discrimination by Label.

Table 1: Examples of posts for each label

| Label | Examples | Ratio (%) |
|---|---|---|
| Gender | My roommate wants her boyfriend to stay overnight in our shared room for a full week. Hi everyone, I'm in a bit of an awkward situation. I'm a first-year student living in a community-style dorm. My roommate has been dating her boyfriend, who goes to a college about three hours away, for six months now. She returned from visiting him over the weekend and unexpectedly told me about her plan. | 25.47 |
| Race | Thanks to her, I'll be confidently graduating this May with a BS in Computer Science. Lǎoshī—I met my Chinese professor, lǎoshī, last semester. My advisor, Dr. W, suggested I take a foreign language, explaining that learning another language helps train the brain to grasp coding languages more effectively. | 13.71 |
| Profession | I've noticed that some professors lack industry experience in their fields. These are usually those with PhDs but little else. They tend to focus heavily on theory, whereas professors with industry experience approach the course from a more practical perspective. In my experience, the latter is generally more effective. | 34.47 |
| Religion | My parents are deeply religious and want me to attend a Christian college, but I'd prefer a school with a vibrant social scene and parties. Are there any Christian colleges more relaxed about the religious aspect? | 15.15 |
| Gender & Race | College students often disregard norms, like wearing pajamas to the library or sleeping under tables. One even stood naked in a campus bathroom, washing his face while international students looked on, likely confused. Such behavior can unintentionally reinforce cultural stereotypes about U.S. college life. | 4.17 |
| Gender & Profession | How do you handle difficult professors? What can you do when a professor clearly gives higher grades to the "pretty girls" in class instead of those who submit better-quality work? | 6.23 |
| Gender & Religion | Expelling someone for inciting rape: I know an extremist Muslim woman from South Asia who went to the UK to study bar-at-law. | 7.99 |
| Race & Profession | Would you find it offensive to read a novel in your English class that includes the "n" word? In 2020, is it appropriate for college literature courses to cover works containing the "n" word? Does it make a difference if the author is African American or a southern white writer? | 4.03 |
| Race & Religion | I'm passionate about studying interior design at Chaminade University in Honolulu, but as an openly gay man, I'm concerned about whether I'd feel comfortable at a religiously Christian school. Does the religious affiliation significantly impact the campus environment, or is it more of a formality? | 7.12 |
| Profession & Religion | Should your minor align closely with your major and career aspirations, or should it reflect a personal interest? What choice did you make? For instance, consider a Criminology major pairing a minor in Psychology versus one in a subject like Judaic Studies. | 7.13 |

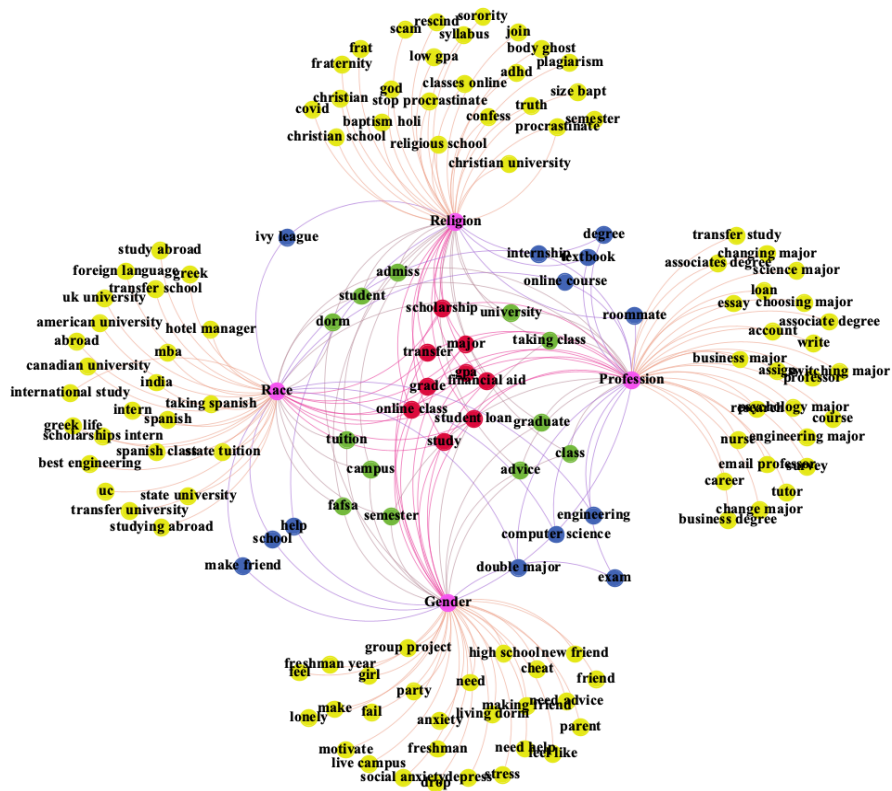

Figure 3: Stereotype Keyword Network. The red nodes at the center represent keywords that appear in all four labeled subgroups. Green nodes indicate keywords associated with three labeled subgroups, blue nodes represent keywords linked to two labeled subgroups, while yellow nodes are related to each subgroup and red nodes are associated with four labeled subgroups. This graph convention is applied to Figure 4 and 5.

Figure 4: Separation Keyword Network. Nodes near the center of the graph represent keywords that frequently appear when more minority groups overlap.

Figure 5: Discrimination Keyword Network. Graph for the emergence or overlap of one or more minority statuses.